\documentclass[a4paper,12pt]{article}
\let\TW=\textwidth

\def\BC{\begin{center}}
\def\EC{\end{center}}
\usepackage{graphicx}

\def\si{${\sf S}_{\sf 1}$}
\def\sii{${\sf S}_{\sf 2}$}

\title{Phase Diagram of the Two-Dimensional\\ Complex Ginzburg-Landau Equation}

\author{{\Large Hugues Chat\'e$^{\rm a,b}$ and Paul Manneville$^{\rm b,a}$}\\
$^{\rm a}$ CEA --- Service de Physique de l'Etat Condens\'e,\\
Centre d'Etudes de Saclay, 91191 Gif-sur-Yvette, France\\
$^{\rm b}$ LadHyX --- Laboratoire d'Hydrodynamique,\\
Ecole Polytechnique, 91128 Palaiseau, France}
\date{}

\begin{document}

\maketitle

\noindent This plain \LaTeX\ version of the article, originally published as:\\
\centerline{Physica A {\bf 224} (1996) 348--368}\\ 
is placed on arXiv because the publisher's link:\\
\centerline{http://www.sciencedirect.com/science/article/pii/0378437195003614}\\
only provides badly resolved black-and-white scans of the figures in place of the initial more readable color pictures.

\begin{abstract}
After a brief introduction to the complex Ginzburg-Landau equation,
some of its important features in two space dimensions are reviewed.
A comprehensive study of the various phases observed numerically
in large systems over  the whole parameter space is then presented.
The nature of the transitions between these phases is investigated
and some theoretical problems linked to the phase diagram are discussed.
\end{abstract}

The complex Ginzburg-Landau equation (CGL) is one of the most
important simple nonlinear partial differential equations for two main
reasons. First, as we will briefly recall below, it
arises as the natural description of many physical situations,  or at
least is the ``kernel'' of many systems of amplitude equations.
Second, its solutions display a very rich spectrum of dynamical
behavior when its parameters are varied, reflecting the interplay
of dissipation, dispersion and nonlinearity. 

Here, we give a brief and mostly qualitative report of the various
regimes observed in the two-dimensional case, and discuss several theoretical
aspects of these numerical findings.

\section{Introduction to CGL}

A large body of work has already been devoted to the CGL equation, which reads:
\begin{equation}
\label{cgl}
\partial_t A = A + (1+i b_1) \nabla^2 A - (b_3 - i) |A|^2 A
\end{equation}
where $A$ is a complex field.

In the context of amplitude equations~\cite{AMPLI,PCHMCC}, 
which are large-scale descriptions
of physical systems passed (and near) symmetry-breaking instability thresholds,
the CGL equation
has been recognized as the relevant equation for the slow modulations
of oscillations in a continuous medium near a Hopf 
bifurcation~\cite{CGL-GEN}. More generally, it appears in the description
of spatially-extended systems when oscillations or waves are present.

\begin{figure}
\BC
\includegraphics[width=0.65\TW]{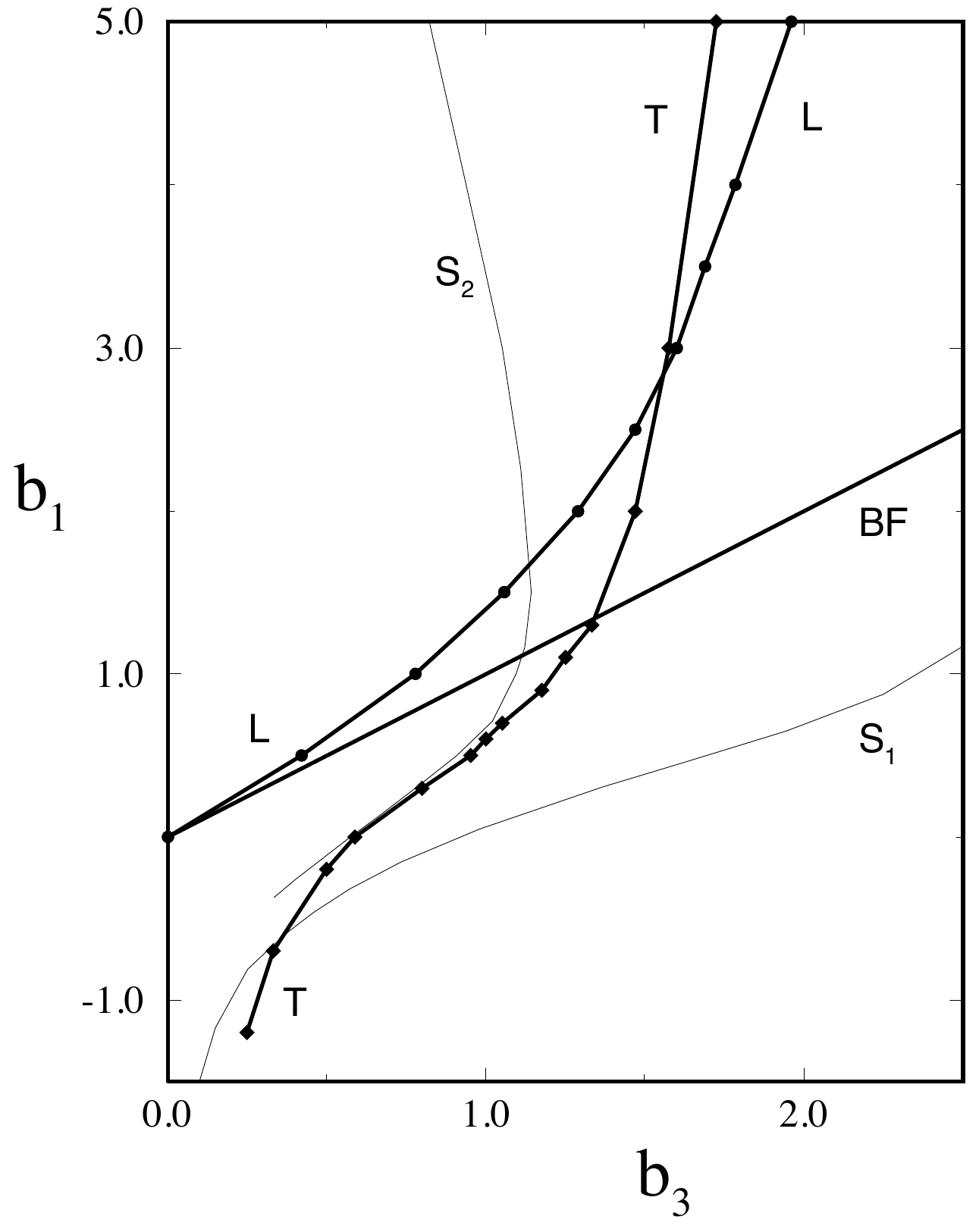}
\EC
\caption{Phase diagram of the two-dimensional CGL equation. 
Phase turbulence is observed between lines {\sf L} and {\sf BF},
defect turbulence to the left of line {\sf T}, and frozen states exist 
(approximately) to the right of line \sii.
Details are given in the text.}
\end{figure}

Under the form (\ref{cgl}), the CGL equation has been reduced (without loss of generality)
to its simplest from, with only two real parameters, $b_1$ and $b_3$.
The first term of the rhs is related to the linear instability mechanism
which led to  oscillations. The second term accounts for diffusion and
dispersion, while the cubic nonlinear term insures ---if $b_3>0$, otherwise
other terms may be necessary--- the saturation  of the linear instability
and is involved in the renormalisation of the oscillation frequency.
Two important limits are worth mentioning: when $b_1=0, 
b_3 \rightarrow \infty$, one has the real Ginzburg-Landau equation,
which possesses a Lyapunov functional and thus exhibits only relaxational
dynamics~\cite{RGL}. 
When $b_1 \rightarrow \infty, b_3=0$ dispersion plays the essential
role, as one recovers the nonlinear Schr\"odinger equation~\cite{NLS}.
In the general case, sustained spatio-temporally disordered regimes are 
observed in large regions of the parameter plane (Fig.~1).

The genericity  of the CGL equation, associated to its relative simplicity,
has made it one of the favorite playgrounds for testing ideas about 
spatiotemporal chaos in a rather realistic context~\cite{PCHMCC}. 
It is only recently, 
though, that a comprehensive study has been undertaken, as it was realized 
that away from the intricacy of the bifurcation diagrams at small sizes,
there exists a crossover size beyond which chaos becomes extensive
and can be characterized by intensive quantities independent of system size,
boundary conditions, and, to a large extent, 
initial conditions~\cite{HSG-EXTEN}.
Indeed, when chaos is extensive, statistical approaches are legitimate and 
should provide rather simple descriptions.
In this context, bifurcation diagrams ---typically used for small dynamical
systems--- are replaced by ``phase diagrams'' delimiting the regions
of  different statistical signatures in parameter space.

Whereas the one-dimensional case is now rather well 
known~\cite{CGL1D,NONLIN,SANTAFE}, 
the situation
in two dimensions is much less satisfactory, mainly because previous work 
was devoted to punctual problems rather than to acquiring a global picture
of the properties of the equation. Here, thanks to current computer power,
we provide a comprehensive overview of the 
two-dimensional CGL equation in the large-size limit~\cite{NOTE1}.

\section{Waves, phase instability and defects}

We now introduce a few important features of the CGL equation before 
proceeding to the description of the phase diagram.

Early work on CGL 
has dealt with the problem of the linear stability
of its family of plane-wave solutions 
$A=a_k \exp i(kx+\omega_k t)$
with $a_{k}^2=(1-k^2)/b_3$ and $\omega_k=1/b_3-(b_1+1/b_3)k^2$.
All these solutions are unstable for $b_1>b_3$ (Newell criterion), 
a condition which defines the so-called ``Benjamin-Feir'' ({\sf BF}) 
line (Fig.~1).
For $b_1<b_3$, plane-wave solutions with 
$k^2<k_{\rm max}^2 = (b_3-b_1)/(3b_3-b_1+2/b_3)$ are linearly
stable~\cite{ECKHAUS}.  The instability of 
the travelling wave solutions above the {\sf BF} line is readily verified
as to be  linked to the ``gauge'' invariance of the equation, i.e.
its invariance by an arbitrary phase shift ($A\rightarrow A\exp i\phi_0$).
Near the {\sf BF} line, the amplitude modes are strongly damped,
``slaved'' to the marginal phase mode, so that one
often speaks of a {\it phase instability}. 
This instability 
has been conjectured to lead to a disordered regime called 
{\it phase turbulence}~\cite{PHASTURB}, 
in which the field $A$ never reaches zero, so that
the phase $\phi=\arg A$ is defined everywhere.
Near the {\sf BF} line, the phase gradient $\nabla \phi$
is expected to remain small, and a systematic expansion can be performed,
leading to a description of the large-scale dynamics in terms of the 
phase only.
The behavior of the resulting series of {\it phase equations} obtained
by truncation of this expansion is
 actually not very well known, except
for the Kuramoto-Sivashinsky (KS) equation to which these equations 
reduce infinitely 
close to the {\sf BF} line. If phase turbulence (i.e. spatiotemporal
chaos) has been established for the 
KS equation, it is the subject of an ongoing controversy away from the 
{\sf BF} line, both for the phase equations and for the CGL equation 
itself~\cite{PHASTURB,SANTAFE,CONTRO-PHAS,HSG-L1}.

Another important feature of the CGL equation is 
the structure, nature, and role of
``defects'', i.e. points in space-time where $A=0$. At such points, the phase
is not defined, and it varies by a multiple of $2\pi$ 
when going around them.
For space dimensions $d\ge 2$, defects are topologically constrained.
This has been recognized 
for a long time as one of the salient features of the two-dimensional
CGL~\cite{COULLET,SPIRAL}. 
For $d=2$, defects are points and can only  appear and disappear 
by pairs. For small enough $b_1$ values, they
appear in two different types, ``spirals'' and ``shock-line vertices''
(Fig.~2). The shock-line vertices have mostly been considered as ``passive''
objects which play no important role. However accurate this statement may be,
it remains that 
the spiral defects have attracted the most
attention~\cite{SPIRAL,KRAMER-SPI}.
In spite of all these efforts, no exact expression is available; on
the other hand, much is known about the core structure and the ``wings'',
i.e. the emitted outward-going
waves. Away from the core, these waves are asymptotic to a 
planewave
with a well-defined wavenumber $k_{\rm sp}$ 
depending only on $b_1$ and
$b_3$. The stability properties of the $k=k_{\rm sp}$ planewave solution give rise
to two important lines in the $(b_1,b_3)$ parameter plane.
On the \si~ line, $k_{\rm sp}=k_{\rm max}$, the maximum wavenumber
of linearly stable planewaves (Fig.~1). This line
delimits the region of linear stability of the $k=k_{\rm sp}$ wave.
To its left, the wave is linearly unstable (perturbations grow exponentially
in phase space); this is in fact a {\it convective} instability: a 
{\it localized} perturbation indeed grows, but is advected away from its 
initial position at the group velocity of the $k=k_{\rm sp}$ planewave.
At this initial position, the solution relaxes to the planewave.
According to Aranson et al.~\cite{KRAMER-SPI}, 
the $k=k_{\rm sp}$ planewave becomes 
{\it absolutely} unstable to the left of 
the \sii~ line: any initial perturbation
grows at its initial location (in addition to spreading in the direction
of the wave). 

\begin{figure}
\includegraphics[width=0.48\TW]{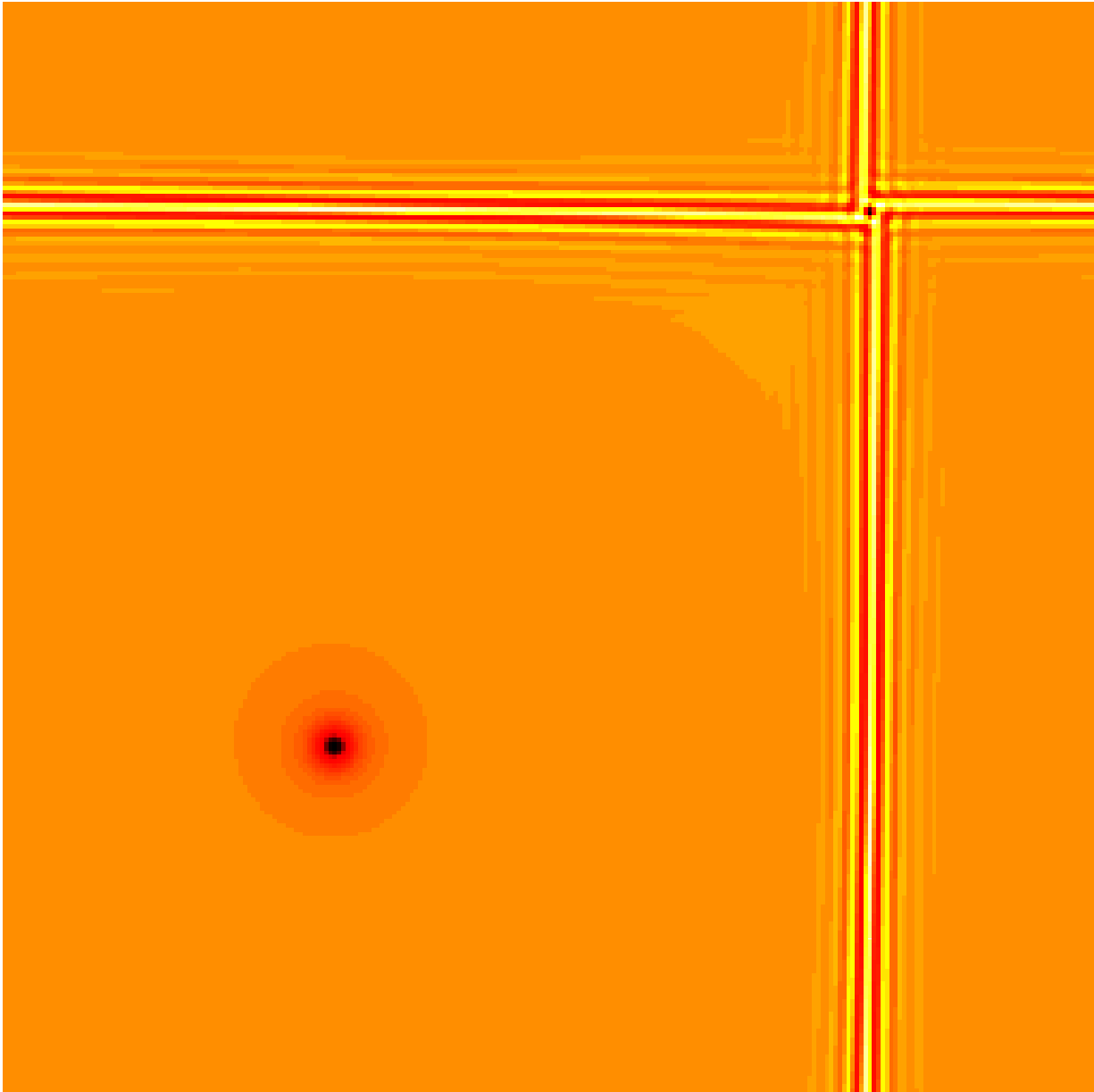}\hfill
\includegraphics[width=0.48\TW]{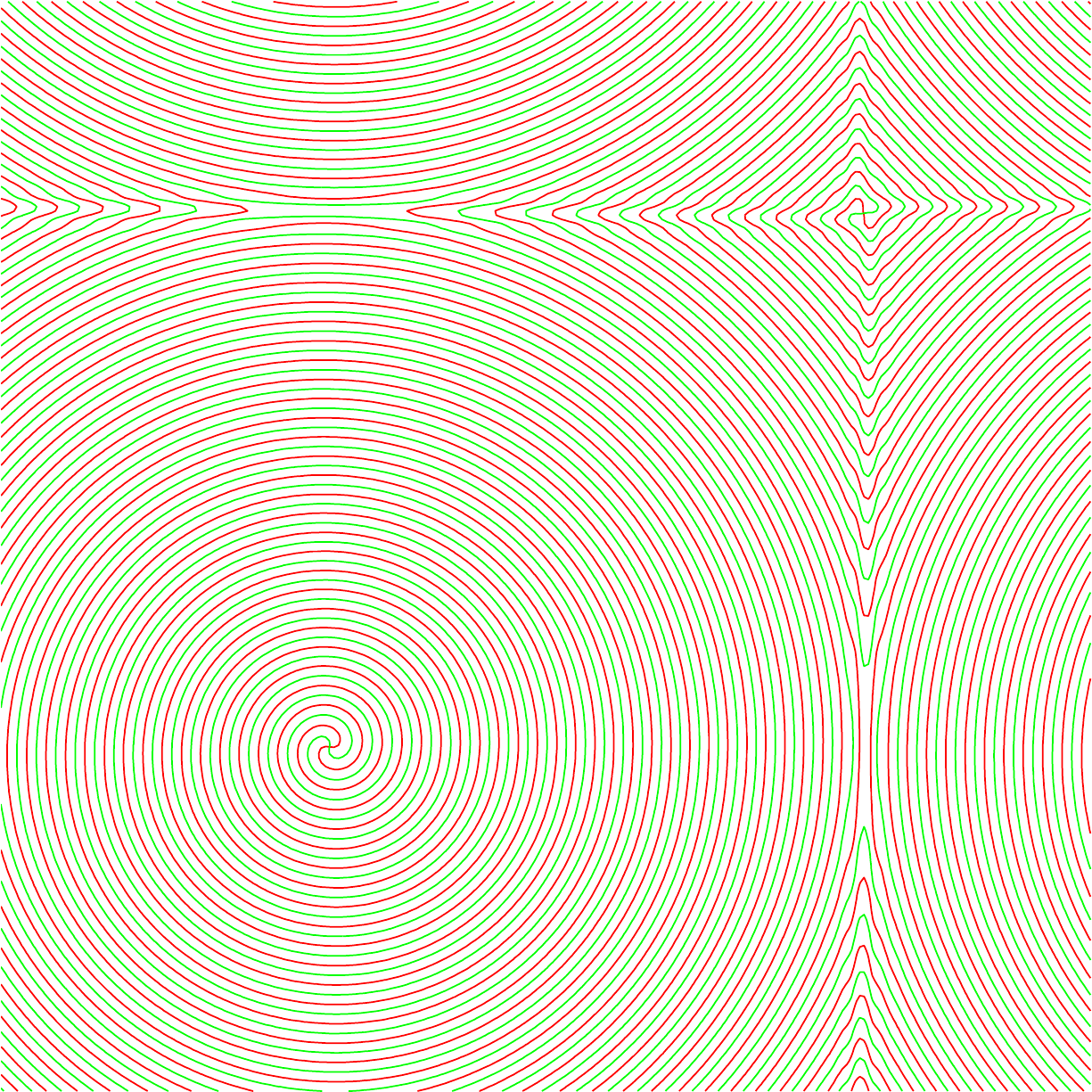}
\caption{Snapshot of a simple frozen configuration with one spiral defect
and one shock-line defect. System of size $L=512$ with periodic boundary 
conditions and parameters $b_1=2$ and $b_3=1.67$. 
(a): image of $|A|$ in color scale from $|A|=0$ (dark red) to $|A|=1.33$ 
(light yellow);
(b): lines ${\rm Re}(A)=0$ (green) and
 ${\rm Im}(A)=0$ (red). }
\end{figure}

It is not exactly known how these stability limits of the  planewave
with wavenumber $k_{\rm sp}$
are related to the actual stability properties of the spiral solution
and to its observability in an experimental context. In most of the region 
of the parameter plane of considered here, the spiral  solution exists and is 
core-stable~\cite{KRAMER-CORE}. 
In consequence, it can be argued that its 
stability properties are essentially related to those
of the asymptotic $k=k_{\rm sp}$ planewave in a {\it semi-infinite} domain
(with the core sitting at one end). We now formulate, at a somewhat conjectural
level, the stability properties of the spiral solution and their consequences
observable in experiments (be they numerical or not).
To the right of line \si, one expects the spiral to resist a (small) 
amount of noise, due to its ``complete'' (core and wings) linear stability.
Between \si~ and \sii, perturbations are amplified but convected away from 
the core at the group velocity of the $k=k_{\rm sp}$ solution. 
Numerical experiments
have shown that the spiral is most sensitive to perturbations in the crossover
region between the core and the wings~\cite{TBP}. 
At a given level of (experimental) 
noise, perturbations coming from this region are  the most dangerous ones.
This convective instability in fact
takes the form of growing oscillations of the {\it modulus} $|A|$ as 
one goes away from the core (see Fig.~6). 
Experimentally, these oscillations do not saturate, and the 
wave breaks down, creating more defects. 
This mechanism defines a maximum radius $R_{\rm noise}$ which limits the size
of observable spirals, and depends on the instability rate and 
(weakly) on the noise level. 
Approaching the limit of absolute 
instability (line \sii), this diameter goes to zero, and beyond \sii~
the spiral is ``completely'' unstable and cannot be observed in an
experimental (noisy) context.

Finally, we note that if defects do play an important role 
in the
two-dimensional CGL equation, as we will show below, their topological
character is not crucial in determining the dynamical regimes: 
for $d=1$, localized ``quasi-defects''
---where $|A|$ remains locally very close to zero---
have been shown to be the key-ingredient in some disordered 
regimes~\cite{NONLIN}.
But the topological constraint on defects for $d\ge 2$ does provide them 
with a large domain of existence in parameter space, insuring their
relevance in most of the regimes of interest.

\section{Phases}

Fig.~1 shows the phase diagram of the two-dimensional CGL equation
established from a numerical exploration of systems of linear size
of the order of $L=512$ with periodic boundary conditions, using
a pseudospectral code. Details about the integration scheme and the numerical
protocol will be given elsewhere~\cite{TBP}. The various transition lines
are discussed in detail in Section~4.

As in the one-dimensional case~\cite{CGL1D,NONLIN}, 
two types of disordered regimes can be 
distinguished, depending on whether they exhibit defects or not. To the right
 of the line {\sf L} in the parameter plane, {\it phase turbulence} 
(no defects)
is observed, whereas {\it defect turbulence} 
occurs to the left of line {\sf T}.

\subsection{Phase turbulence}

Between the {\sf BF} and the {\sf L} lines, spatiotemporally chaotic regimes
of phase turbulence ---where no defect occurs--- are observed. With periodic 
boundary conditions, the total phase gradient across the system 
(the ``winding number'') is conserved.
This introduces a new invariant to the problem. Most results reported
here (in particular the location of line {\sf L}) are for the case
of zero winding number.

\begin{figure}
\includegraphics[width=0.48\TW]{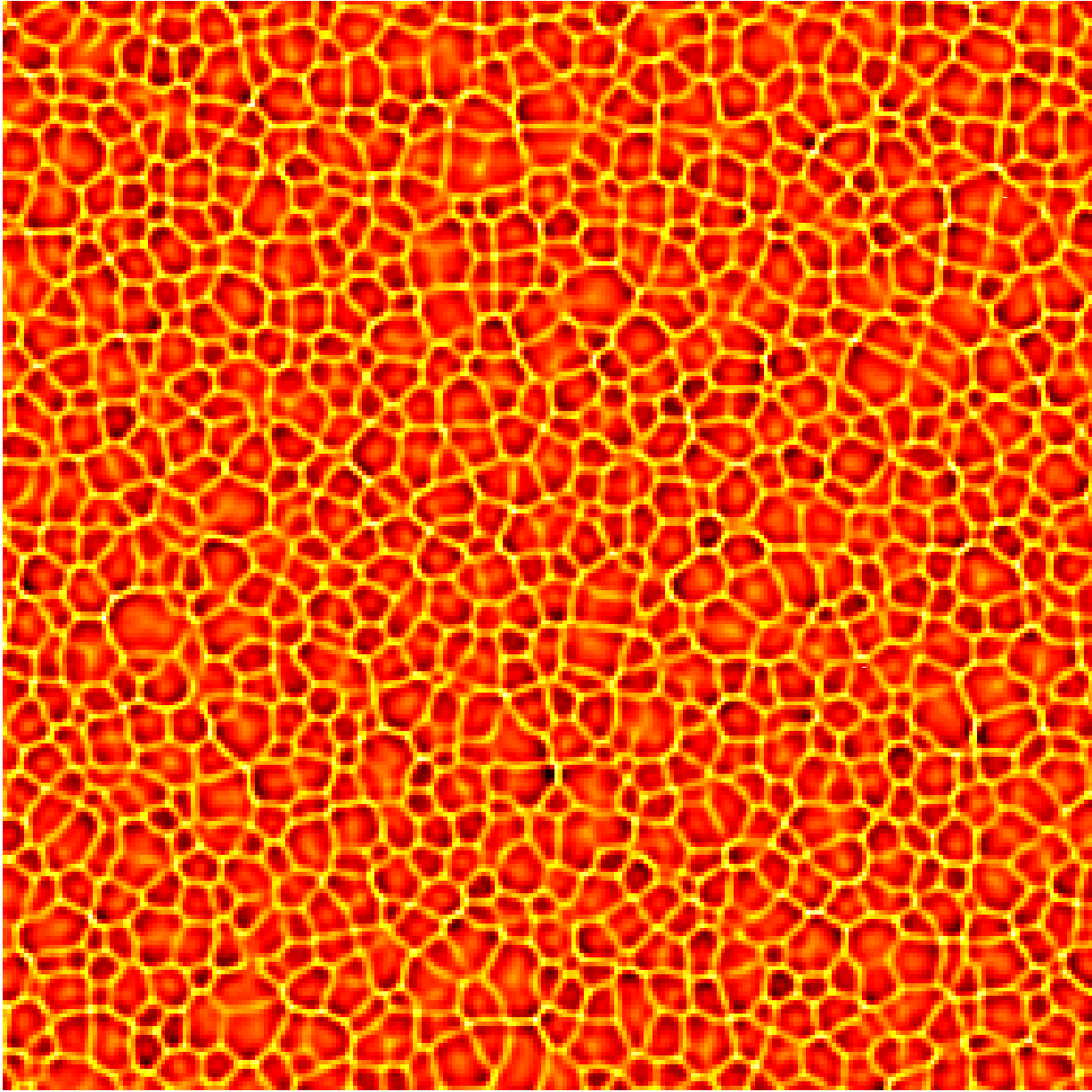}\hfill
\includegraphics[width=0.48\TW]{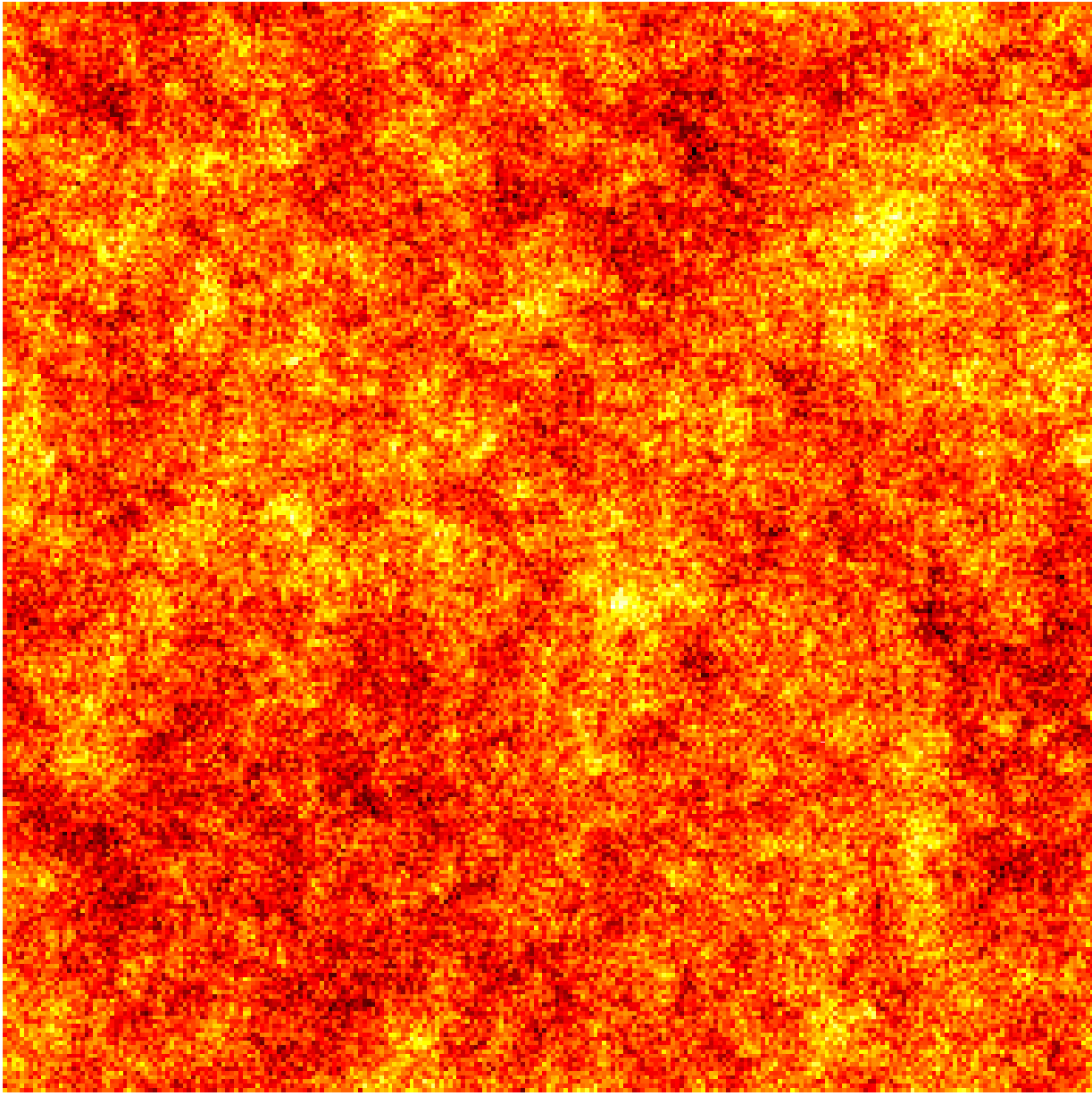}
\caption{Snapshot of phase turbulence in a system of linear size $L=5120$
with periodic boundary conditions and parameters $b_1=2$ and
$b_3=1.33$. (a): field $|A|$ in a sub-system of linear size 
$\ell=640$; color scale
from $|A|=0.87$ (dark red) to $|A|=1.12$ (light yellow); (b): 
phase field $\phi=\arg A$ in the whole system (color scale
from dark red to light yellow over the total range of variation of the phase
$\Delta \phi \sim 4.27$).}
\end{figure}

In phase turbulence, the solution consists of a disordered cellular structure,
(best seen in plots of $|A|$ or $\nabla \phi$)
slowly evolving in time (Fig.~3a). 
The typical size of the cells diverges like $(b_3-b_1)^{-1/2}$ 
when approaching the {\sf BF}
line; this size is in fact of the order of the wavelength of the most 
unstable mode in the corresponding KS equation.
If the correlations of the modulus $|A|$ still decay rapidly, 
those of the phase $\phi$ decay slowly, with power-law-like behavior.
This is apparent in the large-scale modulations of the phase field
(Fig.~3b). In Section~5, we discuss the effective large-scale model
for phase turbulence and the asymptotic behavior of the correlations
in phase turbulence.

Even though they are certainly important to better understanding the dynamics
of phase turbulence, the ``elementary processes'' at play in this regime are
not known. For $d=1$, it has been shown that propagative structures are the 
objects triggering the breakdown of phase turbulence~\cite{SANTAFE}. 
Here, no equivalent 
has so far been found (see below). One should also investigate whether 
the evolution of the cellular structure involves some of the elementary 
events observed in the coarsening of soap froths~\cite{FROTH} 
(even though these cannot 
account for all the dynamics here, 
since the cellular structure is statistically
stationary). Knowledge of the local dynamics 
is necessary to build a large-scale effective description of two-dimensional
phase turbulence in the spirit of the work of Chow and Hwa~\cite{HWA} for the 
one-dimensional KS equation.

\subsection{Defect turbulence}

Defect turbulence is the most chaotic regime of the two-dimensional
CGL equation: correlations decay exponentially,  with short correlation
lengths and times. Depending mostly on $b_1$, the space-time signature
of the solutions varies. For large $b_1$, the density of defects is large,
they come and go rapidly, and they rarely form spirals (Fig.~4).
Indeed, it can be argued that defects {\it per se} are not crucial features
in this case. ``Amplitude turbulence'' is a better name for such
spatiotemporal chaos regimes. Furthermore, increasing $b_1$ toward the 
nonlinear Schr\"odinger equation limit, {\it pulses} become the relevant
objects: the solutions consist of localized regions where $A\neq 0$.
Approaching the {\sf BF} line, the defect density decreases, the characteristic
scales increase, and spirals can be observed.  In fact well-developed
spirals can only be observed, in the defect turbulence region, to the right
of the \sii~ line (see the discussion of the transition lines in the 
next section).

\begin{figure}
\includegraphics[width=0.48\TW]{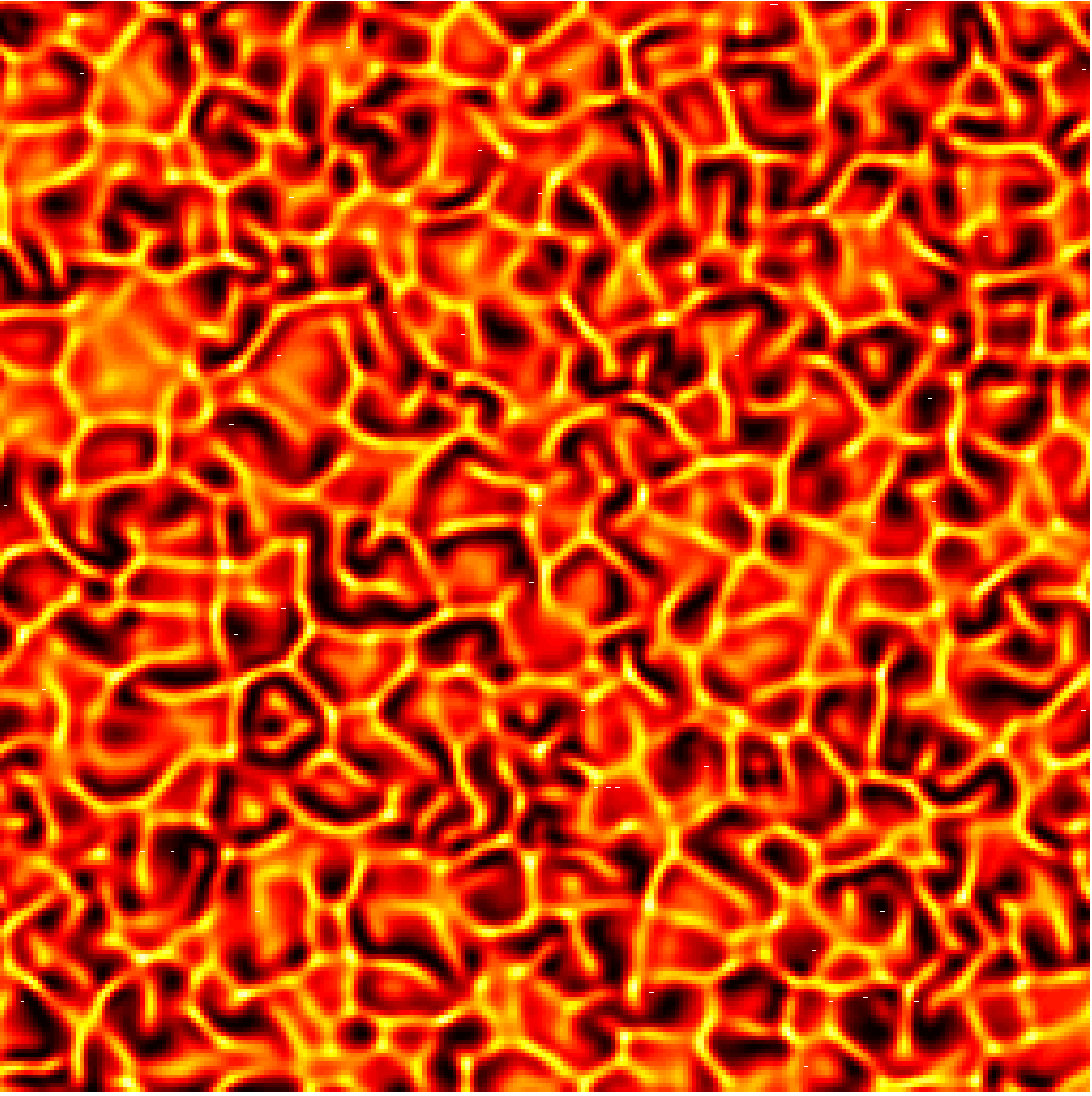}\hfill
\includegraphics[width=0.48\TW]{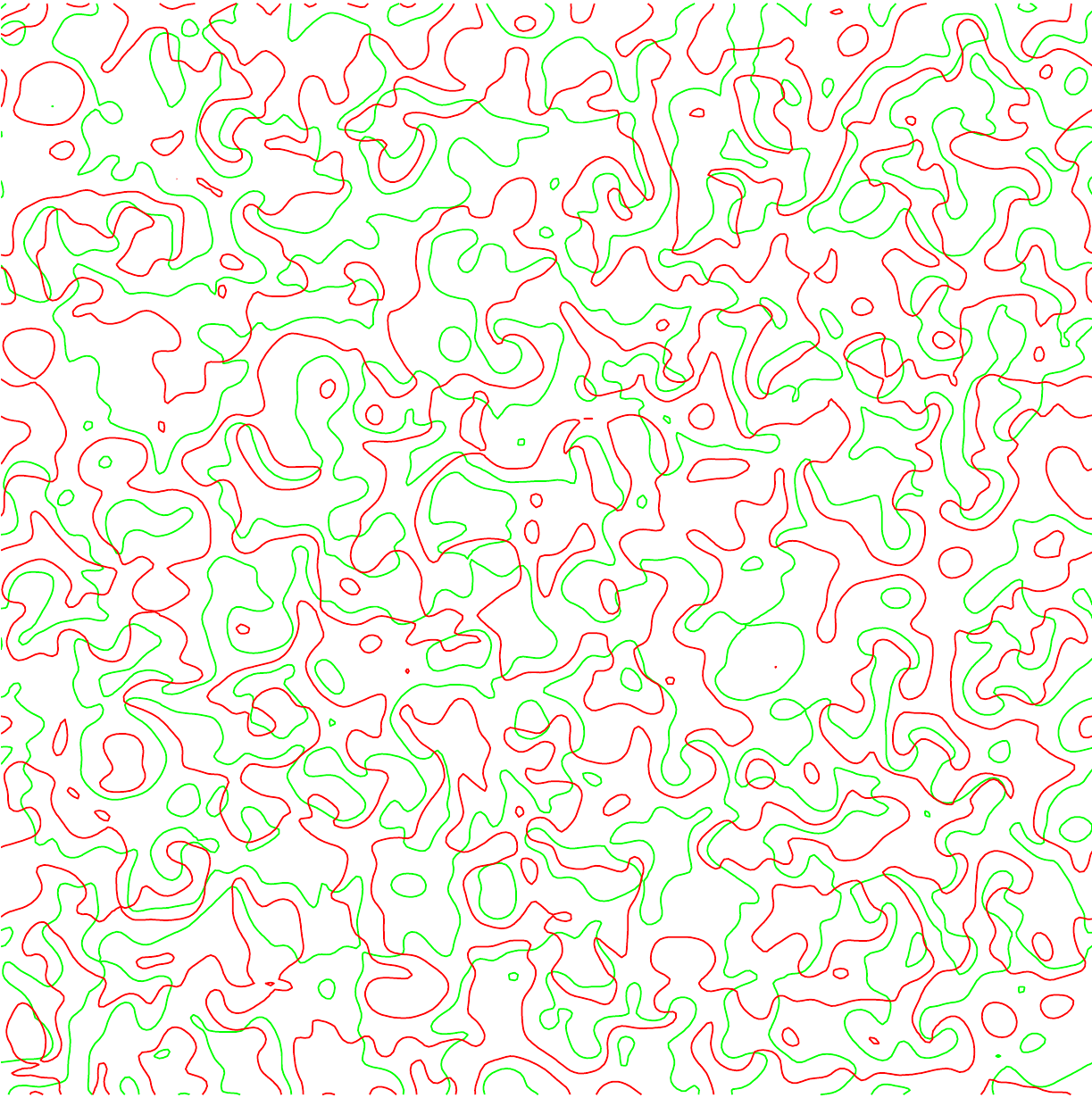}
\vspace{4ex}

\includegraphics[width=0.48\TW]{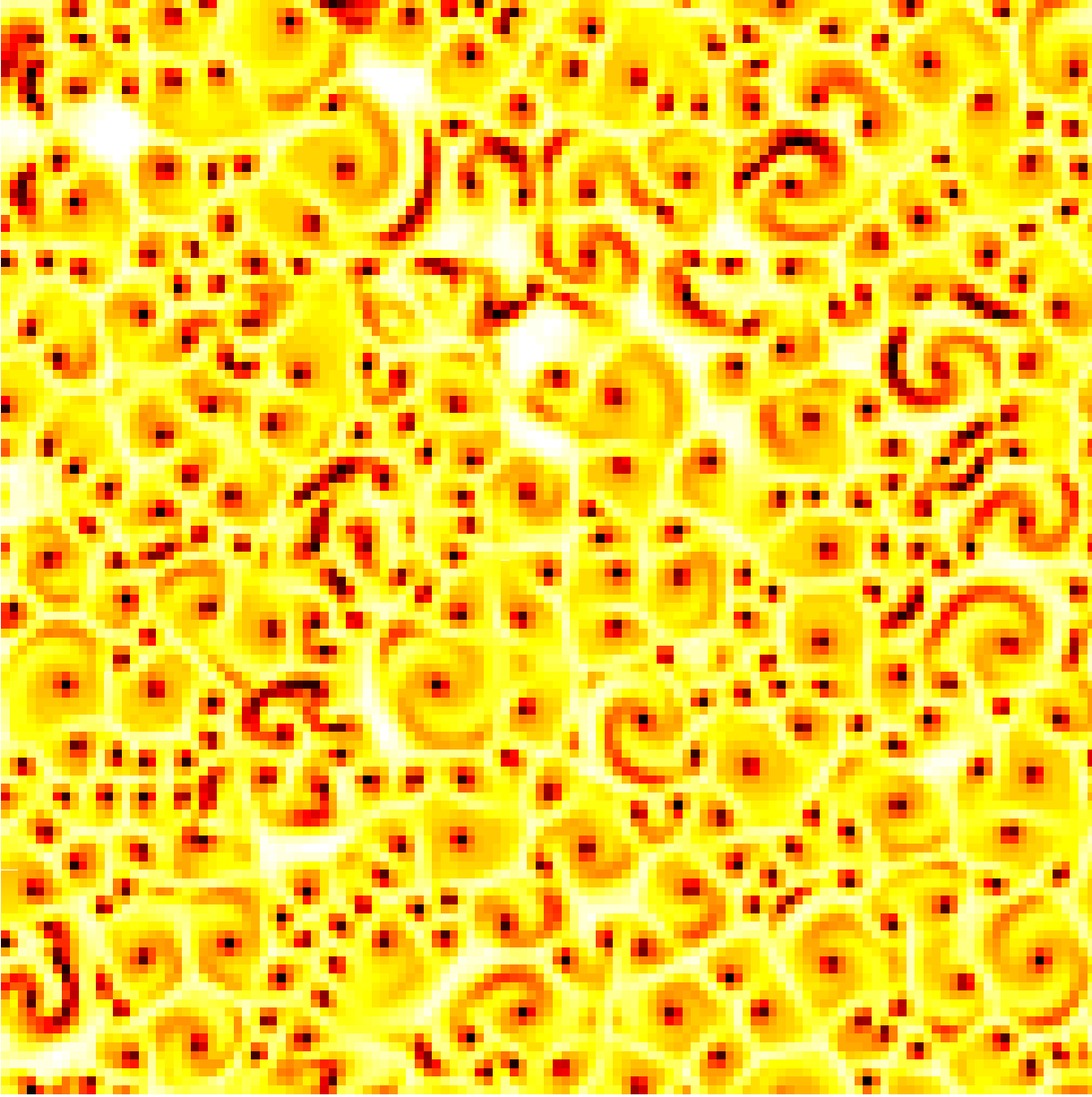}\hfill
\includegraphics[width=0.48\TW]{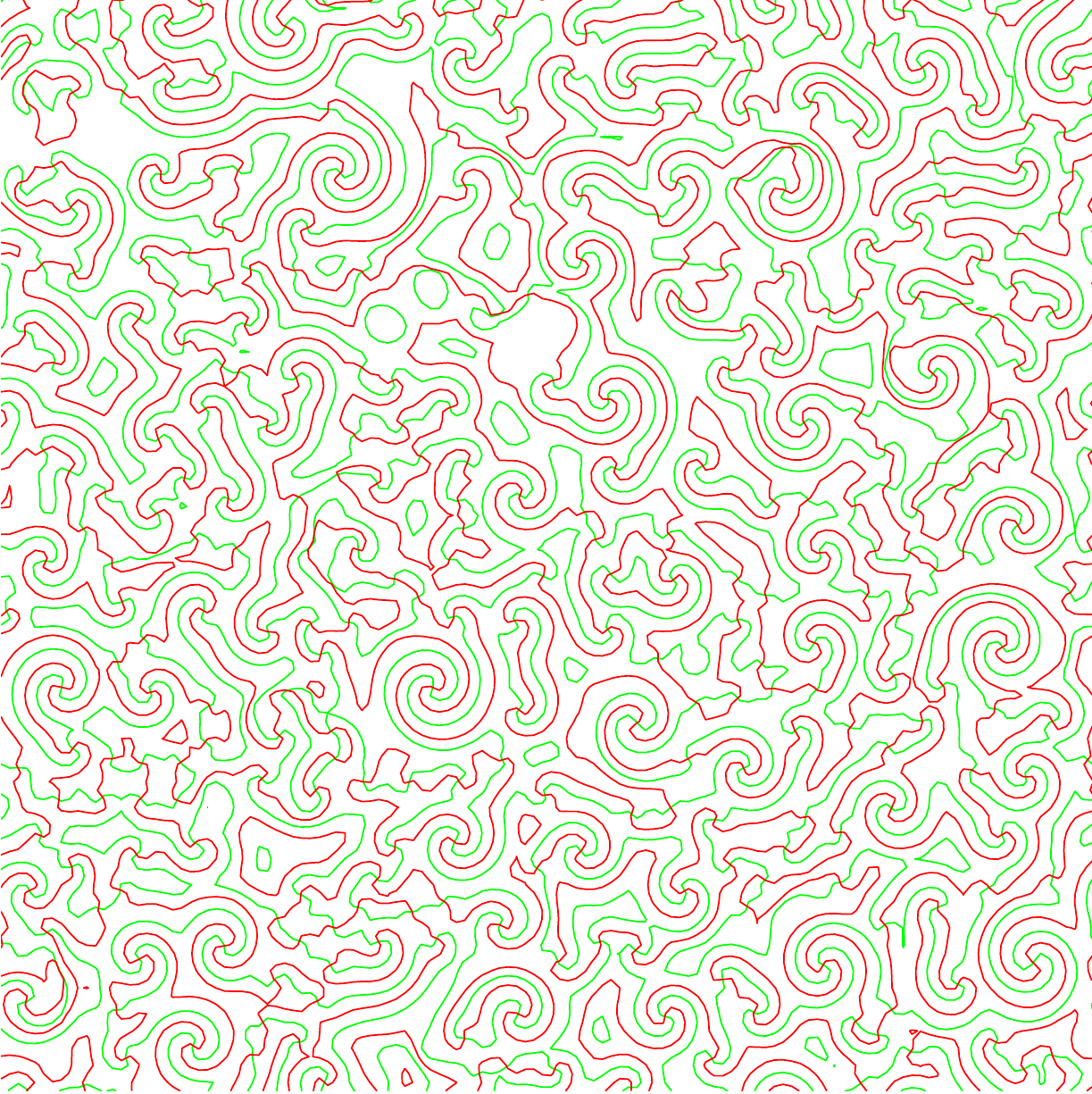}\hfill
\caption{Snapshots of defect turbulence in a system of linear size $L=256$
with parameters $(b_1,b_3)=(2,1)$ (a-b) and $(b_1,b_3)=(0,0.56)$ (c-d).
There are 268 defects in the first case, but no
well-formed spirals are observed; in the second case, on the other hand,
spirals are clearly visible.
(a): field $|A|$; color scale from $|A|=0$ (dark red) to $|A|=1.32$ 
(light yellow); 
(b): ${\rm Re}(A)=0$ (green) and
 ${\rm Im}(A)=0$ (red) lines; 
(c): field $|A|$; color scale from $|A|=0$ (dark red) to $|A|=1.0$ 
(light yellow); 
(d): ${\rm Re}(A)=0$ (green) and
 ${\rm Im}(A)=0$ (red) lines.}
\end{figure}

\subsection{Frozen states}

Cellular structures also appear in the two-dimensional CGL equation
in the form of quasi-frozen arrangements of spiral defects surrounded 
by shock lines. In these states, the field $|A|$ is generally completely
stationary in time.
The network of these lines 
form the cells of these spatially-disordered states (Fig.~5). 
Non-spiral defects lie at the shock-line vertices, sometimes also along the 
shock lines themselves, in metastable arrangements. 
Because the timescales involved are very long, it is actually 
difficult to decide when these structures stop evolving. 
Residual, intermittent, local rearrangements
--- less and less frequent along time--- are observed, 
and this relaxation process is reminiscent of that
taking place in
glasses. In fact, much remains to be done in order to decide to what extent
these dynamical states are glassy states. A first study along these lines
can be found in Huber~\cite{HUBER}. A particular point of interest is to
investigate whether some kind of aging phenomena are taking place
in these frozen structures.

The frozen states are easily observed in the region of the parameter space
to the right of line {\sf T}, where they are the only asymptotic solutions
possessing defects. Their total domain of existence in the $(b_1,b_3)$ plane
can be estimated on the basis of the stability properties of spirals,
as discussed in Section~2. The frozen states do not exist to the left of line
\sii, since there the spirals are absolutely unstable. On the other hand,
nothing precludes their existence to the right of \sii. 
The size of the cells is not limited, except in the presence of noise, since
in this case, between \sii~ and \si, spirals have a maximum radius
$R_{\rm noise}$. In practice, the dynamical ``history'' which led
to a given frozen structure greatly influences the distribution of 
sizes of cells in the structure (see the discussion in section~4.1 below).

\begin{figure}
\includegraphics[width=0.48\TW]{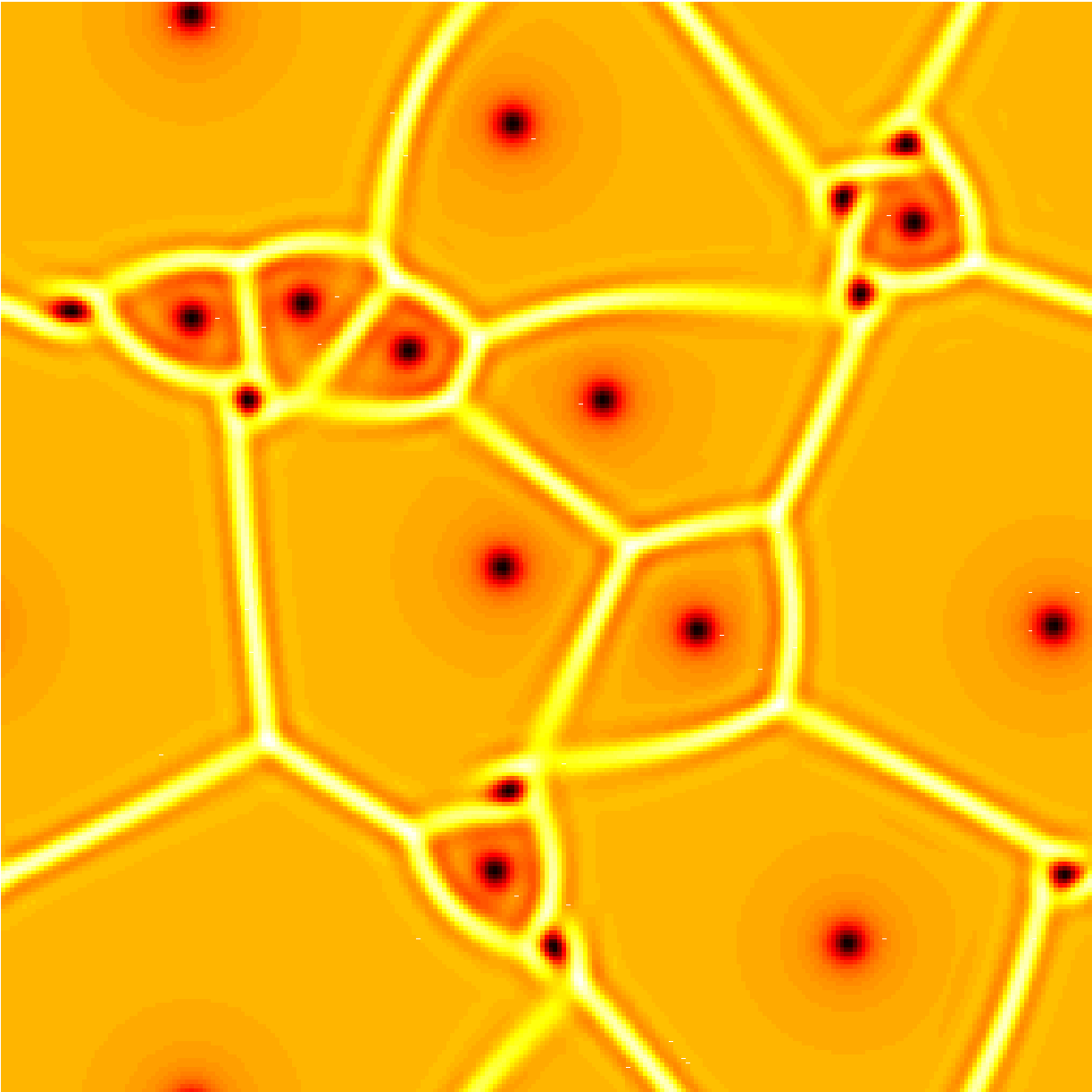}\hfill
\includegraphics[width=0.48\TW]{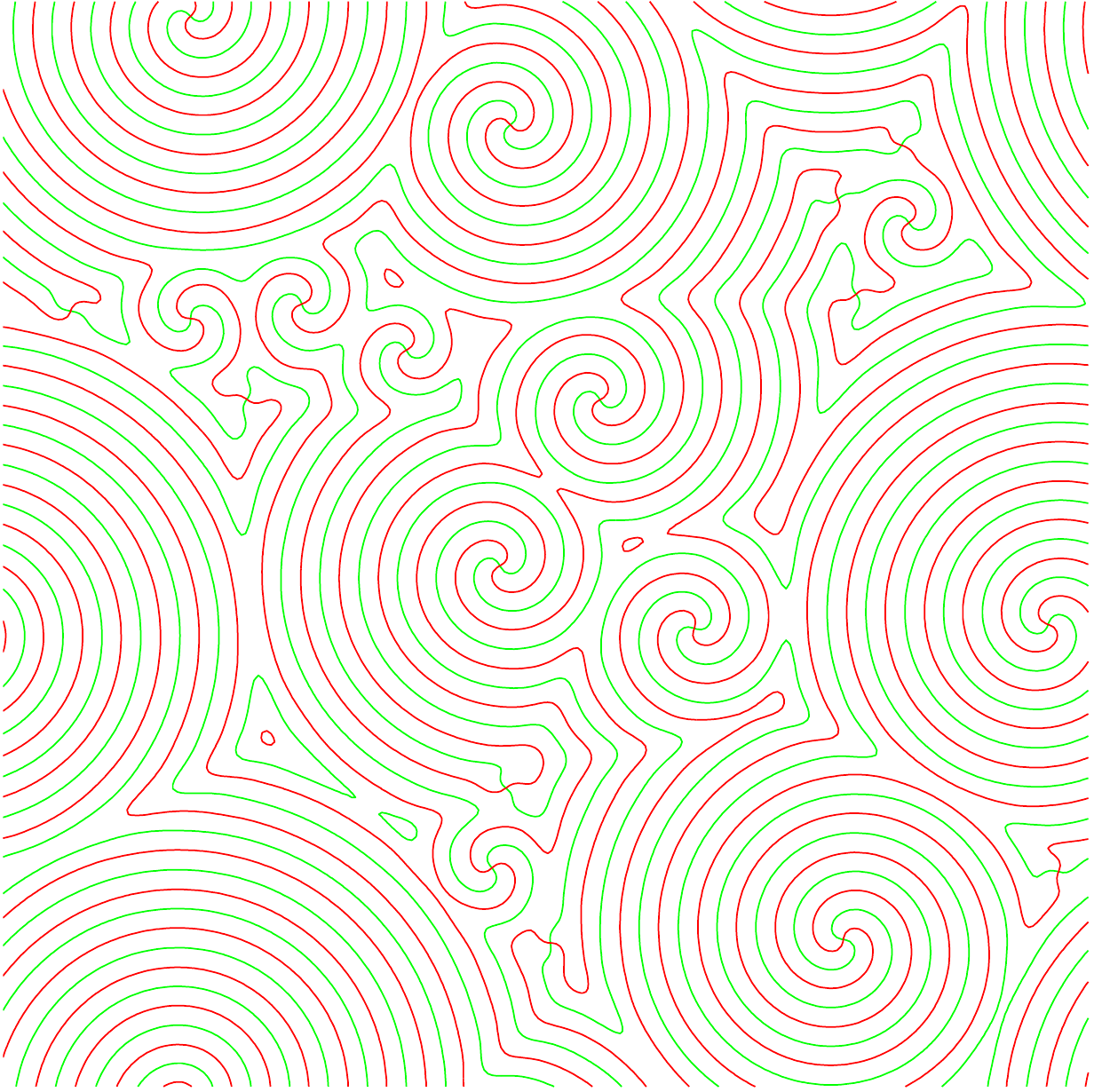}
\caption{Frozen state in a system of linear size $L=256$ with parameters
$b_1=2$ and $b_3=5$.
(a): field $|A|$; color scale from $|A|=0$ (dark red) to $|A|=1.17$ 
(light yellow);
(b): lines ${\rm Re}(A)=0$ (green) and ${\rm Re}(A)=0$ (red).}
\end{figure}

The actual observation
 of frozen states in the region between lines \sii~ and {\sf T}, where defect
turbulence exists, is not easy, though, because these states are
metastable with respect to defect turbulence. Coming, for example, from 
a frozen asymptotic state to the right of line {\sf T}, the parameters
have to be changed ``adiabatically'' to prevent the nucleation of defect
turbulence. Even then, the necessary rearrangements of the cellular structure,
which involve the rapid motion of some defects, most often trigger
the ``melting'' of the frozen structure. 
Frozen structures are most easily observed far to the right of line {\sf T},
and especially to the right of \si.
Their domain of existence
probably extends to large values of $b_3$ (except maybe
for large $|b_1|$). At any rate, along the $b_1=0$ axis, it extends to the
real Ginzburg-Landau ($b_3\rightarrow \infty$) limit, where the spirals become
the vortex excitations of the XY model~\cite{CGLXY}.

\section{Transitions}

The respective domains of existence of the three disordered phases described
above are delimited by the lines {\sf BF}, {\sf T}, {\sf L}, and \sii.
We discuss now the nature of the various (phase) transitions observed when 
crossing these lines and comment on the relative stability of the 
disordered phases.

\subsection{Lines {\sf T} and \sii}

As already mentioned, line {\sf T} delimits the (numerically estimated) domain
of existence of {\it sustained} regimes of defect turbulence. Starting from 
a defect turbulence regime, {\it increasing $b_3$}, this highly chaotic regime
is maintained until line {\sf T} is crossed; defect turbulence is only
transient then, and is
followed by the nucleation of a frozen state (Fig.~6). As observed
by Huber et al.~\cite{HUBER}, 
this transition is indeed reminiscent of a first-order
phase transition. Depending both on the amplitude of the ``quench'' beyond
line {\sf T} (i.e. the distance of the current parameters to line {\sf T})
and the $b_1$ value of the crossing point, the duration of the transient varies
widely. The smaller the quench and the larger $b_1$, the longer the transient.
As a matter of fact, line {\sf T} can be seen as the line where this transient
is infinite.

The nucleation process involves the appearance of a sufficiently large spiral
core. 
To the right of line {\sf T}, the spiral may then grow, but its size is limited
to a maximum radius $R_{\rm turb}$ (Fig.~6a).
This radius results from the interaction between the outward-going
spiral waves and the strong, finite-amplitude fluctuations characteristic
of the defect turbulence ``bath'' surrounding it. These fluctuations
trigger the most unstable mode of the spiral solutions, 
i.e. the oscillations of $|A|$ which are the signature of the nonlinear
stage of the convective instability, and influence the spiral wave
{\it inward}. We stress that this is different from the problem 
usually considered when studying convective instabilities. Here a
semi-infinite convectively-unstable medium is put in contact ``downstream''
with a turbulent medium. 
The balance between the (destabilizing) turbulent fluctuations 
and the (regularizing) advection of perturbations by the spiral
waves takes place at the nonlinear level, so that the radius
$R_{\rm turb}$ cannot be determined from the stability properties
of the spiral alone~\cite{TBP}. 

Approaching line {\sf T} from the right, 
$R_{\rm turb}$ decreases. Numerical experiments~\cite{TBP} show that the line
{\sf T'} where $R_{\rm turb}=0$ (not shown in Fig.~1) 
is located to the {\it left} of line {\sf T}.
Note that, 
in contrast with line {\sf T}, line {\sf T'} is defined {\it via} a local 
dynamical phenomenon. In the region between lines {\sf T} and {\sf T'},
``fully developed'' defect turbulence decays to mixed states, i.e.
mostly-frozen structures in which some localized patches of turbulence
subsist (Fig.~6b). Numerically speaking, this residual turbulence does not seem
to vanish at long times; extensive statistical data has to be compiled
in order to decide whether this remains true in the infinite-time limit,
in which case lines {\sf T} and {\sf T'} are distinct. In the other case,
one must conclude that line {\sf T}, in the thermodynamic limit, moves to the
left to coalesce with line {\sf T'}. However, in a similar fashion to what 
happens with line {\sf L} in phase turbulence (see next section),
line {\sf T} is numerically well-defined for all practical
purposes, and it is only a theoretical point to know whether it is distinct
from {\sf T'} in the infinite-time limit. 
We note finally that the frozen states nucleated this way
(i.e. from the spontaneous decay of defect turbulence) possess
a maximal cell size given by $R_{\rm turb}$.

\begin{figure}
\includegraphics[width=0.48\TW]{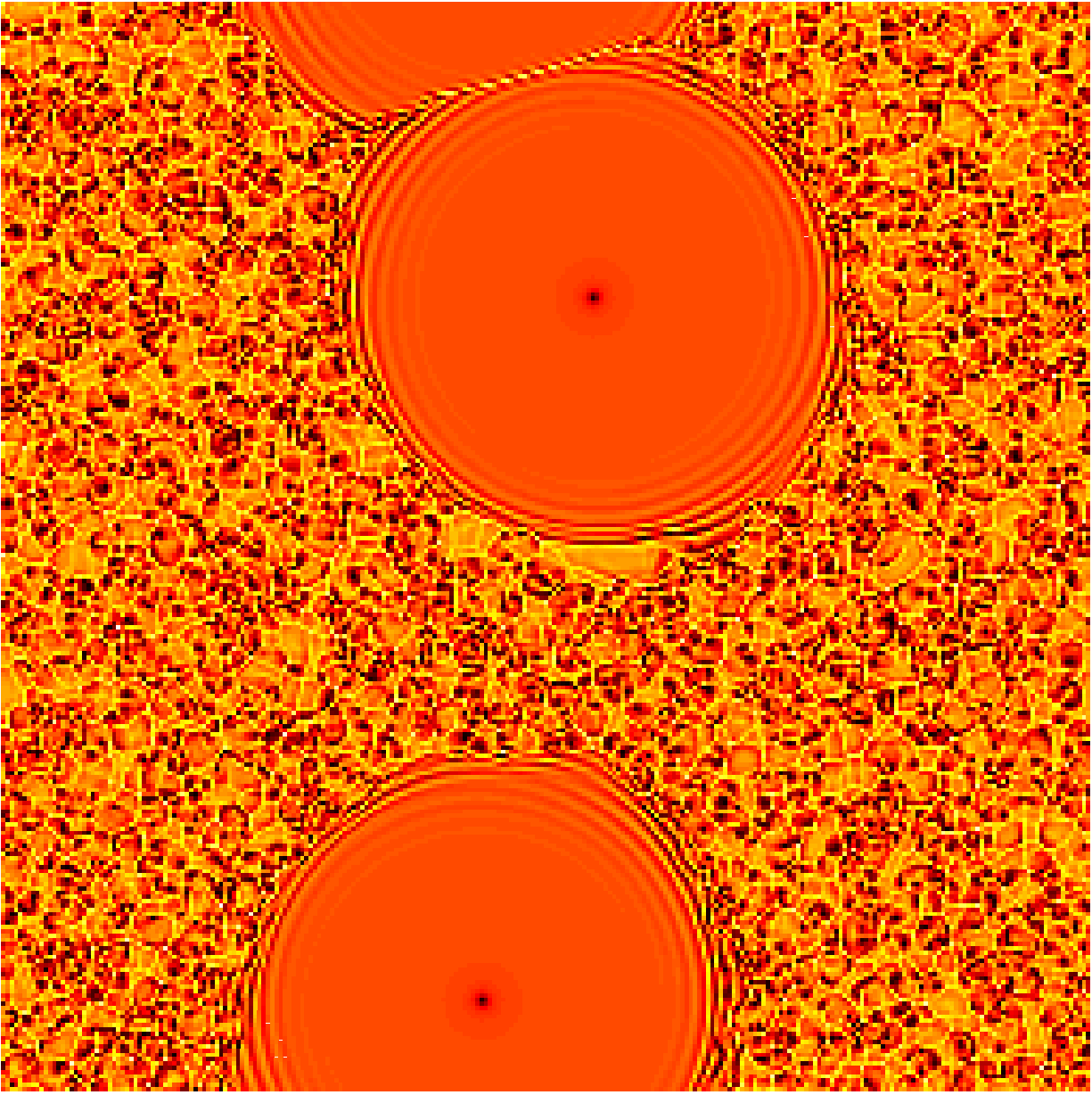}\hfill
\includegraphics[width=0.48\TW]{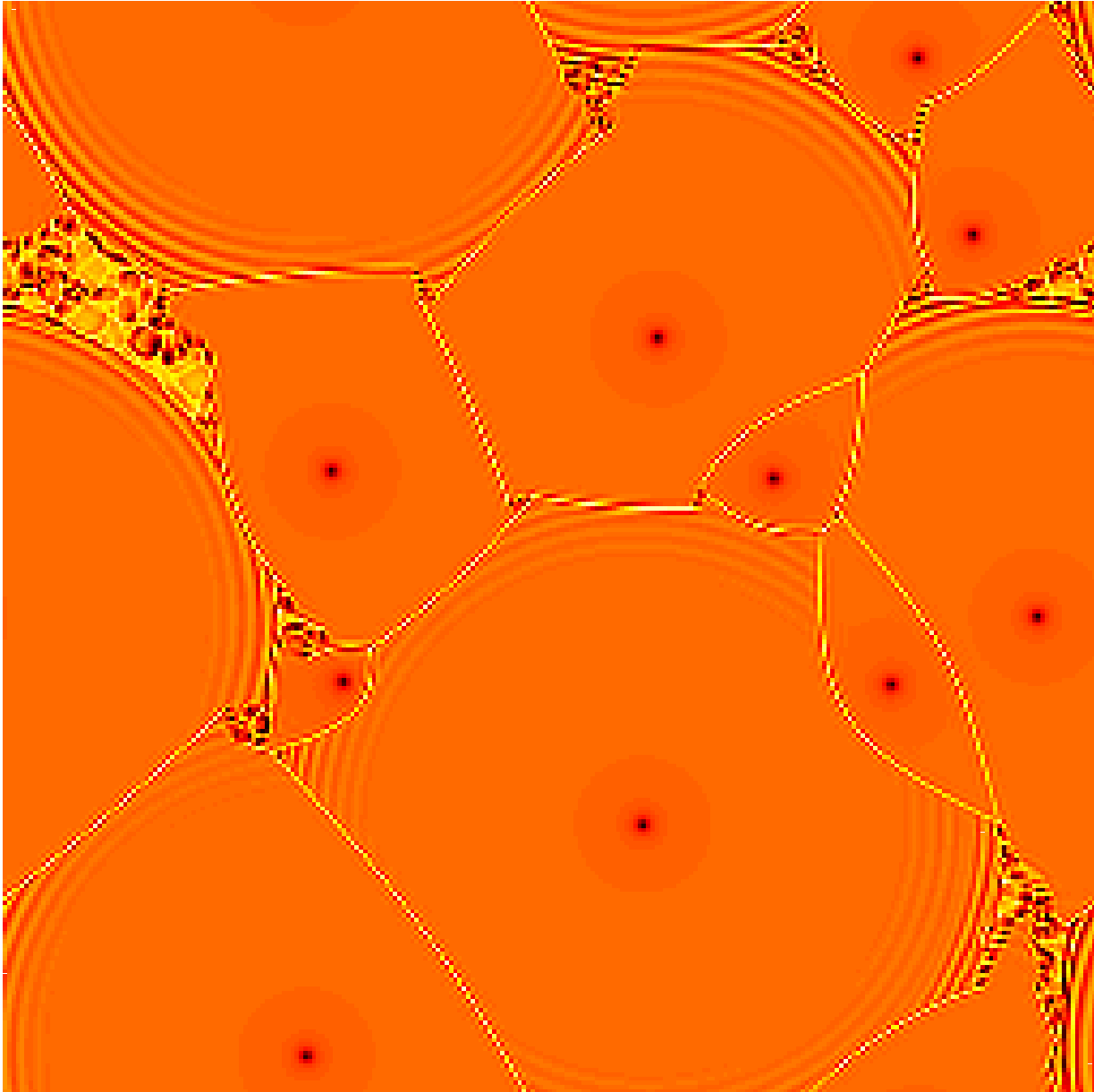}
\caption{Nucleation of a frozen state from defect turbulence in a system
of linear size $L=1024$ with parameters $b_1=2$ and $b_3=1.33$. 
(a): two spirals have been nucleated and have reached their maximal size
$R_{\rm turb}\sim 225$;
(b): asymptotic state consisting of a frozen structure of spirals with
maximum radius $R_{\rm turb}$ with some residual turbulence for 
$L=1024$, $b_1=2$, and $b_3=1.43$. This asymptotic
state is typical of the region between lines {\sf T} and {\sf T'} (compare
with Fig.~5).
(Snapshots of field $|A|$, color scale from $|A|=0$ (dark red) to $|A|=1.22$
(light yellow)). Note the oscillations of $|A|$ away from the spiral cores,
near their maximum radius, which are the signature of the convective 
instability of the $k=k_{\rm sp}$ planewave.}
\end{figure}

The transition from defect turbulence to frozen states is hysteretic: 
coming from a frozen state, and {\it decreasing
$b_3$}, it is, in principle, possible to keep {\it completely}
frozen structures past line 
{\sf T}, (even if ``partially'' frozen structures, 
such as the one shown in Fig.~6b, coexist in the region between lines {\sf T} 
and {\sf T'}). 
As mentioned in Section~3, line \sii~ is only 
an absolute and 
approximate limit of the hysteresis region: in theory, the effects
of the curvature of the waves and the spiral core should be taken into 
account, and in practice, in this
region of parameter space, the frozen
states are easily destroyed by perturbations, so that it is extremely 
difficult to observe the frozen states far to the left of line {\sf T}.
In fact, the real limit of existence of frozen states might be actually
determined by the properties of the {\it nonlinear} stage of 
the convective instability of the waves and possibly also by the stability
properties of the shock-line vertices, the other key-component of frozen 
structures. This very intricate situation will be examined in detail in
\cite{TBP}. The distance
between lines {\sf T} and \sii~ is thus only an approximate measure of the
{\it maximal} width of the hysteresis loop.

\subsection{Line {\sf L}}

Line {\sf L} is the (numerically-determined) line beyond which (to its left)
phase turbulence is only transient. To its right, phase turbulence can be 
observed for as big a system and as long a time as current computers 
allow. A brief discussion of the existence of phase turbulence in the 
infinite-size, infinite-time limit is given in the next section, but, 
numerically speaking, line {\sf L} is rather well defined, 
with the probability of 
breakdown of phase turbulence being finite to its left and essentially zero 
to its right.

\begin{figure}
\includegraphics[width=0.48\TW]{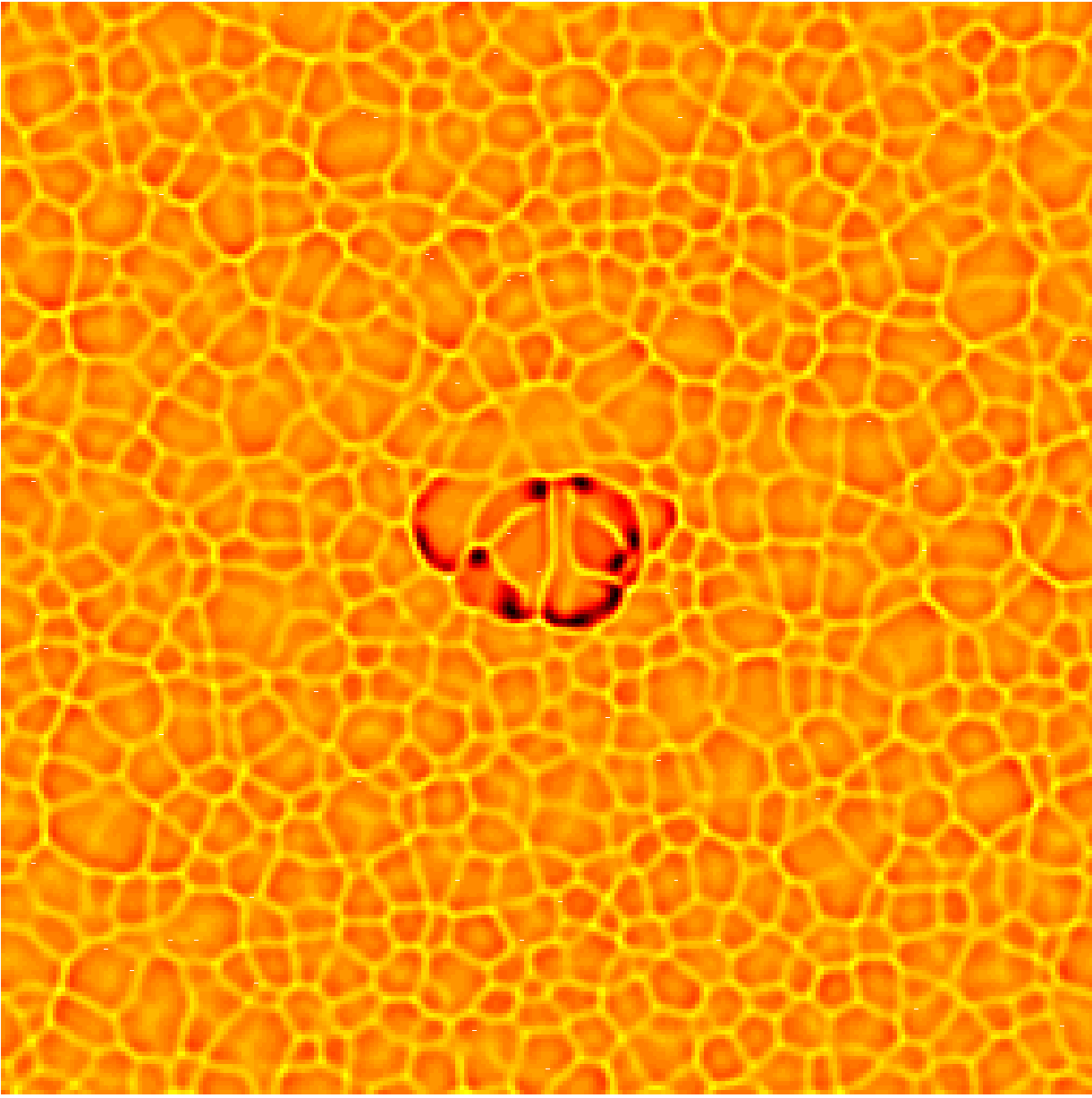}\hfill
\includegraphics[width=0.48\TW]{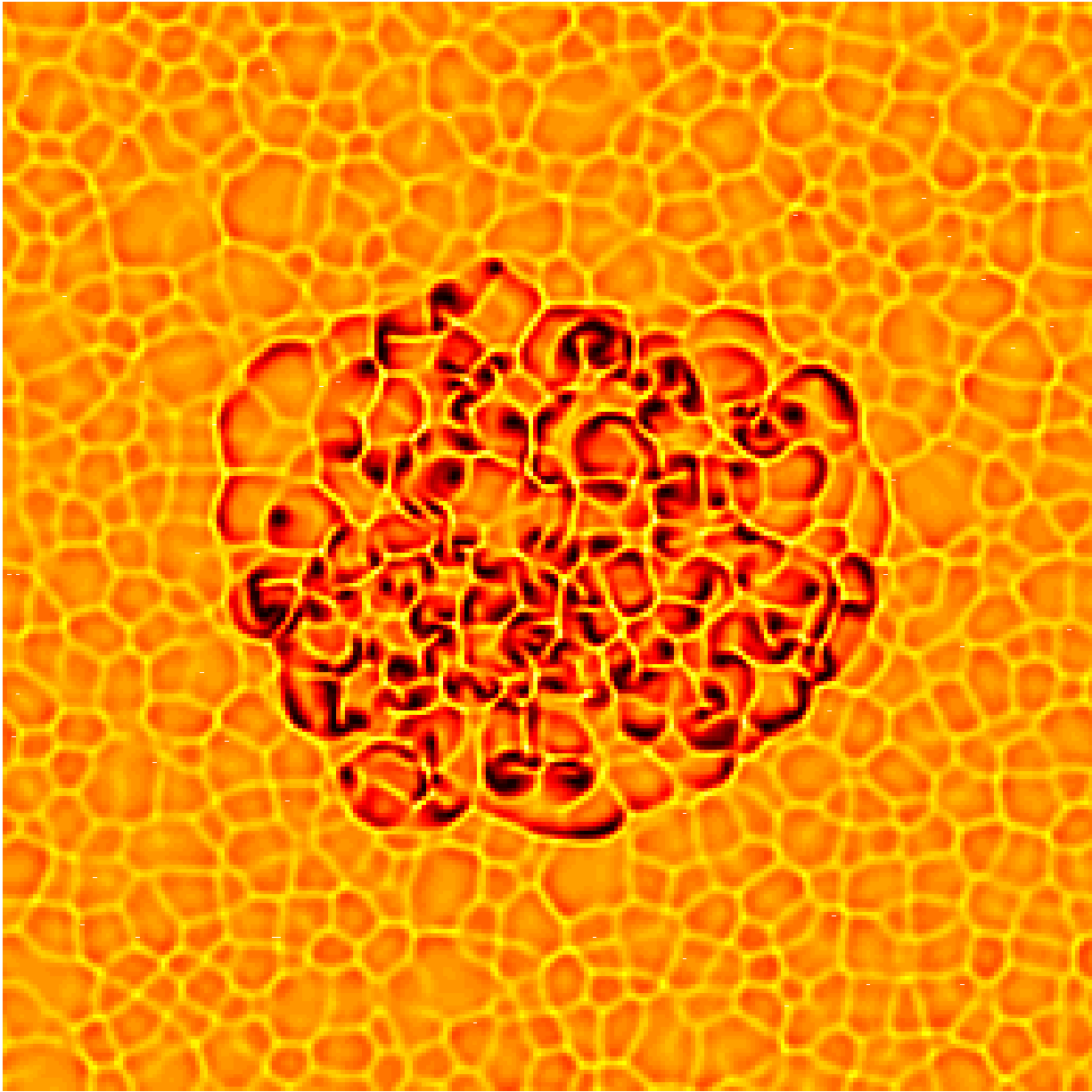}
\vspace{4ex}

\includegraphics[width=0.48\TW]{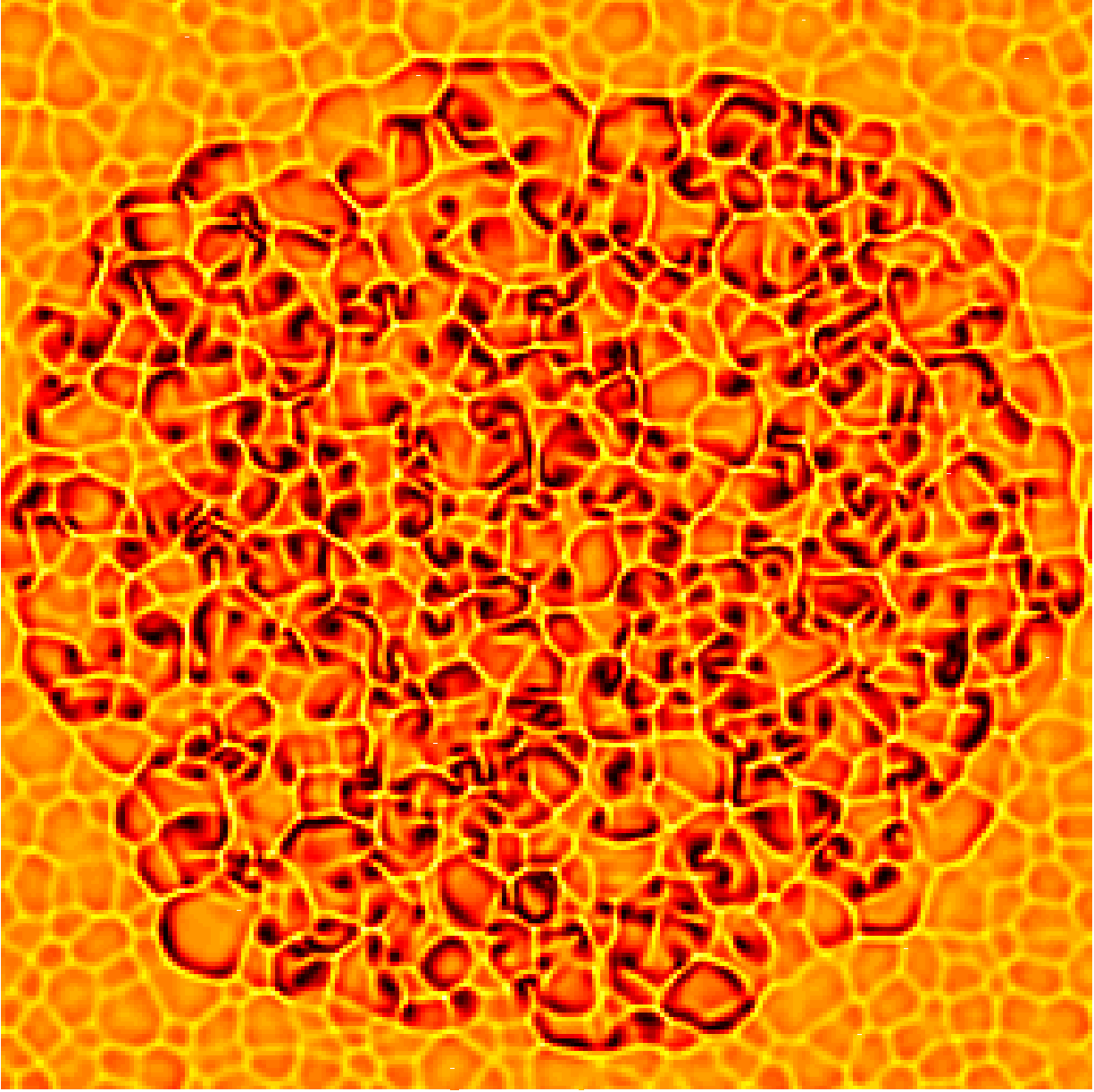}\hfill
\includegraphics[width=0.48\TW]{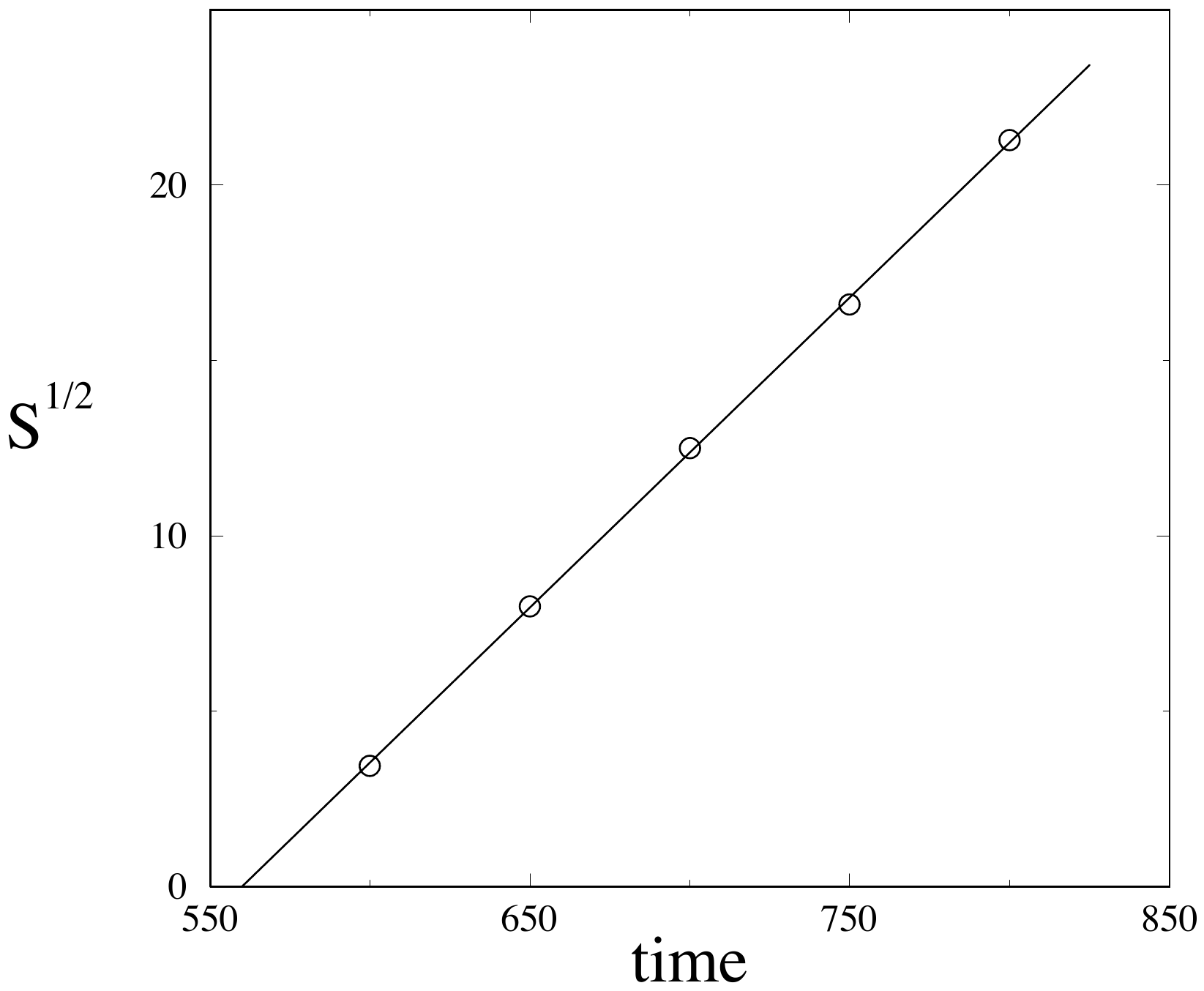}
\caption{Breakdown of phase turbulence in a system of linear size $L=512$
with parameters $b_1=2$ and $b_3=1.28$ coming from a phase turbulent system
at $b_1=2$ and $b_3=1.33$. (a-c): snapshots of field $|A|$ at time $t=600,
700, 800$ in color scale from $|A|=0$ (dark red) to $|A|=1.29$ (light yellow). 
Note the growing ``bubble'' of defect turbulence, whose diameter increases
linearly with time ((d): square root of the surface $S$ 
of the bubble along time).}
\end{figure}

The breakdown of phase turbulence is also a nucleation process (Fig.~7).
A pair of defects is nucleated by some fluctuation on one side of a cell,
 triggering a ``chain-reaction'' leading to the quasi-deterministic
invasion of the system by a growing bubble of the defect turbulence phase.
The diameter of this bubble increases linearly with time (Fig.~7d).

A priori, three sections of line {\sf L} have to be considered, delimited 
by the crossing points with lines {\sf T} and \sii~ (or, rather,
the line actually delimiting the existence of frozen states). 
Below line {\sf T},
the only possible regime is defect turbulence, so that the breakdown
of phase turbulence can only lead to this regime. Between lines
\sii~ and {\sf T}, defect turbulence and frozen states coexist, and the breakdown
could lead to either state. In fact, the nucleation events at the origin of 
the breakdown are always highly chaotic, and only defect turbulence arises
(this is not surprising, considering the metastability of frozen states in 
this region). 
The third portion of line {\sf L}, above its crossing with line {\sf T},
offers an interesting possibility: there, only frozen states are expected 
asymptotically. But the breakdown of phase turbulence first triggers a defect 
turbulence transient which then itself nucleates a frozen state. No direct
transition from phase turbulence to frozen states seems possible, although
it is conceivable that one might observe, in a large system, 
a nucleating frozen state   within 
the growing bubble of defect turbulence invading 
phase turbulence.

Finally, we note, not surprisingly, that the breakdown of phase turbulence
is also an hysteretic transition. Crossing line {\sf L} from left to right,
one remains in either defect turbulence or a frozen state.

\section{Discussion}

After this brief description of the ``phase diagram'', we comment on 
important related points as well as some theoretical problems currently
under investigation.

\subsection{Infinite-size, infinite-time limit}

The phase diagram of Fig.~1 summarizes numerical results. As such, 
even though customary precautions have  been taken (e.g. to insure the 
extensivity of chaos in the disordered regimes), it does not represent the
``true'' phase diagram, i.e. that of the infinite-size, infinite-time,
``thermodynamic'' limit. This question, which also arises in the $d=1$ case,
has been recently investigated~\cite{HSG-L1}, in particular with respect 
to the existence of phase turbulence in the thermodynamic limit (represented
by line {\sf L} here). Indeed, the very existence of phase turbulence is 
questioned. As for the $d=1$ case, it is currently impossible to make a 
definitive statement,
on the basis of numerical simulations alone, as to 
whether line {\sf L} coalesces
with the {\sf BF} line in the thermodynamic limit. 
A careful analysis of statistical data about phase turbulence in one and
two dimensions for various system sizes and integration 
times is under way and will be reported elsewhere~\cite{TBP}.
Extrapolation of size effects, though, seems to  lead to the conclusion
that phase turbulence might not exist in the thermodynamic limit. 
It remains nevertheless that for all practical purposes (numerical or 
experimental), there exists a domain of parameter space where phase turbulence
is statistically stationary and subsists ---even in very large systems--- for 
times as long as desired.
At any rate, Fig.~1 is representative of system sizes and 
integration times
accessible to current computers (say up to linear size L of the 
order of $10^4$ and integration times up to $10^5$), 
and line {\sf L}  
is probably slightly shifted for much larger systems.

The status of line {\sf T}, which delimits the domain of existence
of sustained regimes of defect turbulence, is subjected to similar remarks.
Statistical data about the probability of breakdown (to a frozen state)
should be cumulated in order to estimate the position
of line {\sf T} in the thermodynamic limit. 
This should be completed by 
a detailed study of the variation of $R_{\rm turb}$ with 
parameters $b_1$ and $b_3$ to provide a precise determination of line 
{\sf T'}.
The question, raised in Section~4.1, of whether lines {\sf T} and {\sf T'} 
coalesce in the thermodynamic limit, could thus be addressed.
The existence of 
defect turbulence is not in question, though, as line {\sf T} is bound to be
situated to the right of line \sii~ (which limits absolutely
the domain of existence of frozen states).

\subsection{Cellular structures}

Cellular structures appear both in phase turbulence (Fig.~3) and, of course,
as frozen states (Fig.~5).  In phase
turbulence, the dynamics produces statistically stationary
configurations which can be studied along the lines of what
is usually done for, say, the coarsening of soap froths~\cite{FROTH}.
For example, a first step would be to estimate the statistical properties
of cells ---their sizes, their number of sides--- and a second stage
could consist in determining the local events of which their dynamical
evolution is composed.
As mentioned in section~3.2, 
this should provide a better understanding of the elementary processes
involved in phase turbulence, as well as a better statistical description
of this regime. 
This could also pave the way to a simple ``particle model'' sharing the
same statistical properties, similar to the work of Rost and Krug~\cite{KRUG}
on the KS equation.
In the case of the frozen states, as discussed earlier,
the characteristics of the structures  depend strongly on the ``dynamical
history'' that led to them, so that an
investigation of their geometrical properties must be correlated to
their origin and formation.

\subsection{Complete phase diagram}

Admittedly, the phase diagram of Fig.~1 is not quite complete. 
Work remains to be done, in particular on the three following points:
\begin{itemize}
\item The $b_1 \rightarrow \infty$, ``nonlinear Schr\"odinger'', limit
needs to be clarified: Where are lines \sii~ and {\sf T} located?
How does the {\it core} instability of spirals intervene in the dynamical
regimes?
\item The $b_1 <0$ region, and especially the prolongation of line {\sf T},
should be investigated, as well as the $b_1 \rightarrow -\infty$ limit.
\item It is known that, at least for $d=1$, the CGL equation exhibits
well-behaved disordered regimes in some parts of the $b_3<0$ 
half-plane~\cite{BRETHER},
in spite of the absence of the usual nonlinear saturation mechanism.
Such a possibility should also be explored in the two-dimensional case.
\item the processes  of nucleation of the spiral cores
should be studied in detail, at  least from a statistical point of view.
In particular, the minimal core size that can possibly lead to the growth
of a spiral should be estimated, as well as the probability of
such an event. This would help define a line to the right of line 
{\sf T} beyond which nucleation is immediate (corresponding to the line
``NUC'' in \cite{HUBER}), as well as clarify the nature of the frozen states 
with residual turbulence observed experimentally between lines {\sf T} 
and {\sf T'}.
\end{itemize}

\subsection{Large-scale description}

The possibility of large-scale descriptions of deterministic
spatiotemporal chaos
in terms of Langevin-like, (nonlinear) stochastic equations, is a question
currently being debated, essentially because one can then hope to apply methods
of statistical mechanics~\cite{LANGEVIN}. 
One central point is to investigate to what
extent and under what conditions the local chaotic fluctuations are 
equivalent to a ``noise'' at large scales.

The CGL equation offers, here also, a good testground for such questions.
In this spirit,
phase turbulence 
has been proposed~\cite{GRINSTEIN} to be described, at large scales,
by the noisy Burgers or Kardar-Parisi-Zhang (KPZ)~\cite{KPZ}
equation, which is 
(among other things) a model for the kinetic roughening of stochastic
interfaces. Indeed, since the phase $\phi=\arg A$ is always defined 
in this regime, one can consider the evolution of the phase field only
(at least on an experimental level), which, in turn, can be seen as the 
progression of a $d$-dimensional interface in a $(d+1)$-dimensional medium
(unwinding the phase advance on the real axis). Numerical results for
$d=1$ and $d=2$ seem to confirm the validity of the KPZ ansatz~\cite{TBP}.
The KPZ picture also reveals the asymptotic behavior of 
correlations in phase turbulence. In particular, spatial correlations of 
the phase should decrease either algebraically (``linear regime'' of KPZ)
or like a stretched exponential (``nonlinear regime'').

To go beyond this type of numerical observation, the effective large-scale
stochastic equation has to be built from the original model. An important step
toward this aim has been achieved recently for the one-dimensional
Kuramoto-Sivashinsky equation (which is also believed to be described by KPZ
at large scales~\cite{KSKPZ}).
Carefully studying the elementary mechanisms at the
 origin of spatiotemporal chaos, Chow and Hwa~\cite{HWA} 
have succeeded in calculating,
from data on local chaos only, the parameters of the effective KPZ equation.
It is not clear how such a program could be carried out for the CGL equation
in any of its disordered regimes ---even for phase turbulence---, 
but a detailed analysis of the elementary 
processes at work in each case appears as a necessary step deserving 
further work.

\section{Conclusion}

The general picture of the two-dimensional CGL equation presented here,
even though it should be completed along the lines mentioned above,
already provides a good starting point to
people wanting to study various aspects of spatiotemporal
chaos in this system. In particular, our study should help choose
specific parameter values. It should also help experimentalists
recognize whether the physical or numerical problems they study are typical of
the CGL equation and, if so, of what particular regime. 
Finally, natural extensions of this work include a similar study
of the three-dimensional case, and of the various modifications
of CGL usually considered in the literature.


\begin{thebibliography}{99}
\bibitem{AMPLI} See, e.g., A.C. Newell, ``Envelope Equations'', in
{\it Nonlinear Wave Motion}, Lectures in Applied Mathematics, Vol.~15, 157
(American Mathematical Society, Providence, RI, 1974).


\bibitem{CGL-GEN}  Y.~Kuramoto,
{\it Chemical Oscillations, Waves and Turbulence},
(Springer, Tokyo, 1984);
 J. Lega, ``D\'efauts topologiques associ\'es
\`a la brisure de l'invariance de translation dans le temps,''
Th\`ese de doctorat, Universit\'e de Nice (1989);
W. van Saarloos, ``The Complex Ginzburg-Landau Equation for Beginners'',
in {\it Spatiotemporal Patterns in Nonequilibrium Systems}, P.E. Cladis
and P. Palffy-Muhoray eds., (Addison-Wesley, Reading, 1994).

\bibitem{RGL} See, e.g.: R.~Graham, {\it Phys. Rev. A} {\bf 10} (1974) 1762.

\bibitem{NLS} See, e.g.: A.C.~Newell, {\it Rocky Mountains J. Math.} {\bf 8}
(1978) 25; A.C.~Scott, F.Y.F.~Chu and D.W.~McLaughlin, {\it Proc. IEEE}
(1973) 1443; Y.S. Kivshar and B.A. Malomed, {\it Rev. Mod. Phys.}
{\bf 61} (1989) 762; A.C. Newell, D.A. Rand and D. Russell,
{\it Phys. Lett. A} {\bf 132} (1988) 112;
S. Popp, O. Stiller, I. Aranson and L. Kramer, {\it Physica D} {\bf 84}
(1995) 424.


\bibitem{PCHMCC}  M.C.~Cross and P.C.~Hohenberg, ``Pattern formation
outside of equilibrium'', {\it Rev. Mod. Phys.} {\bf 65} (1993) 851.

\bibitem{HSG-EXTEN} D.A.~Egolf and H.S.~Greenside, 
{\it Nature\/} {\bf 369} (1994) 129.

\bibitem{CGL1D}  B.I. Shraiman, A.~Pumir, W.~van~Saarloos, P.C.~Hohenberg,
H.~Chat\'e, and M.~Holen, {\it Physica~D} {\bf 57} (1992) 241;
A. Pumir, B.I.~Shraiman, W.~van Saarloos, P.C.~Hohenberg, 
H.~Chat\'e, and M.~Holen,
``Phase vs. Defect Turbulence in the One-Dimensional Complex
Ginzburg-Landau Equation,'' in: {\it Ordered and Turbulent Patterns in
Taylor-Couette Flows}, C.D.~Andereck ed.
(New York: Plenum Press, 1992).

\bibitem{NONLIN} H. Chat\'e, {\it Nonlinearity} {\bf 7} (1994) 185.

\bibitem{SANTAFE} H. Chat\'e, ``Disordered Regimes of the One-Dimensional
Complex Ginzburg-Landau Equation'', 
in {\it Spatiotemporal Patterns in Nonequilibrium Systems}, P.E. Cladis
and P. Palffy-Muhoray eds., (Addison-Wesley, Reading, 1994).

\bibitem{NOTE1} A previous study (see \cite{HUBER}), 
using a discretized version of the CGL
equation, attempted such a task.
In view of the results presented here,
it appears that the effects of the discretization scheme on the phase diagram
are rather drastic. In particular, no phase turbulence regime was observed
by these authors, an artifact, we believe, due to their numerical scheme. 
Other discrepancies will be discussed elsewhere~\cite{TBP}.

\bibitem{ECKHAUS} B. Janiaud, A.~Pumir, D.~Bensimon, V.~Croquette,
H.~Richter, and L.~Kramer,
{\it Physica~D} {\bf 55} (1992) 259.

\bibitem{PHASTURB} H. Chat\'e and P. Manneville, ``Phase Turbulence'',
in {\it Turbulence: A Tentative Dictionary}, P. Tabeling and O. Cardoso eds.
(Plenum, New York, 1994) and references therein.

\bibitem{CONTRO-PHAS} H. Sakaguchi,
{\it Prog. Theor. Phys.} {\bf 84} (1990) 792.

\bibitem{HSG-L1}  D.A.~Egolf and H.S.~Greenside, {\it Phys. Rev. Lett.}
{\bf 74} (1995) 1751.

\bibitem{COULLET} Coullet, P., L.~Gil, and J.Lega,
{\it Phys. Rev. Lett.} {\bf 62} (1989) 1619; 
L. Gil, J. Lega and J.L. Meunier, {\it Phys. Rev. A} {\bf 41} (1990) 1138.


\bibitem{SPIRAL} P.S. Hagan, {\it SIAM J. Appl. MAth.} {\bf 42} (1982) 762;
I. Aranson, L. Kramer and A. Weber, ``The Theory of Motion of Spiral
Waves in Oscillatory Media'', 
in {\it Spatiotemporal Patterns in Nonequilibrium Systems}, P.E. Cladis
and P. Palffy-Muhoray eds., (Addison-Wesley, Reading, 1994); {\it Phys.
Rev. E} {\bf 47} (1993) 3231.

\bibitem{KRAMER-SPI} I. Aranson, L. Aranson and L. Kramer, 
{\it Phys. Rev. A.} {\bf 46} (1992) R2992; 

\bibitem{KRAMER-CORE} I. Aranson. L. Kramer and A. Weber,
 {\it Phys. Rev. Lett.} {\bf 72} (1994) 2316.

\bibitem{TBP} H. Chat\'e and P. Manneville, ``Defect Turbulence and Frozen 
States in the Two-Dimensional Complex Ginzburg-Landau Equation'' and 
``Phase Turbulence in the
Complex Ginzburg-Landau Equation'', to be published.

\bibitem{FROTH} B. Levitan and E. Domany,
``Topological model of soap froth evolution with deterministic
T2-processes,'' preprint (1995).

\bibitem{HWA} C.C.~Chow and T.~Hwa, {\it Physica D\/} {\bf 84} (1995) 494.

\bibitem{HUBER} G. Huber, P. Alstr{\o}m and T. Bohr, {\it Phys. Rev. Lett.}
{\bf 69} (1992) 2380; G. Huber, ``Vortex Solids and Vortex Liquids
in a Complex Ginzburg-Landau Equation'', 
in {\it Spatiotemporal Patterns in Nonequilibrium Systems}, P.E. Cladis
and P. Palffy-Muhoray eds., (Addison-Wesley, Reading, 1994).

\bibitem{CGLXY} H. Chat\'e and L-H. Tang, unpublished.

\bibitem{KRUG} M. Rost and J. Krug, {\it Physica D} 88 (1995) 1.

\bibitem{BRETHER} C.S. Bretherton and E.A. Spiegel, {\it Phys. Lett. A} 
{\bf 96} (1983) 152.

\bibitem{LANGEVIN}  See \cite{KSKPZ} and
the discussions in: M.S.~Bourzutschky and M.C.~Cross,
{\em Chaos\/} {\bf 2} (1992) 173;
J.~Miller and D.A.~Huse, {\em Phys. Rev. E.\/} {\bf 48} (1993) 2528;

\bibitem{GRINSTEIN}  G. Grinstein, C. Jayaprakash, and R. Pandit,
``Conjectures about phase turbulence in the complex
Ginzburg--Landau equation,'' to appear in {\it Physica~D}.

\bibitem{KPZ}  M.~Kardar, G.~Parisi and Y.-C.~Zhang,
{\em Phys. Rev. Lett.\/} {\bf  56} (1986) 889;
see also the reviews: J.~Krug and H.~Spohn, Kinetic roughening
of growing surfaces, in: C.~Godr\`eche ed.,
{\em  Solids Far From Equilibrium\/}, (Cambridge University Press, 1991);
T.~Halpin-Healy and Y.C.~Zhang, {\em Phys. Rep.\/}, to appear. 


\bibitem{KSKPZ}  V.~Yakhot, {\it Phys. Rev. A} {\bf 24} (1981) 642;
S. Zaleski, {\it Physica D} {\bf 34} (1989) 427;
K. Sneppen, J. Krug, M.H. Jensen, C. Jayaprakash, and T. Bohr, 
{\it Phys. Rev. A} {\bf 46} (1992) 7351;
I. Procaccia, M.H. Jensen, V.S. L'vov,
K. Sneppen, and R. Zeitak, 
{\it Phys. Rev. A} {\bf 46} (1992) 3220;
V.S. L'vov and I. Procaccia, 
{\it Phys. Rev. Lett.} {\bf69}(1992) 3543;
C. Jayaprakash, F. Hayot, and R. Pandit,
{\it Phys. Rev. Lett.} {\bf71} (1993) 12.



\end{thebibliography}
\end{document}